\documentclass[lettersize,journal]{IEEEtran}
\IEEEoverridecommandlockouts
\usepackage{cite}
\usepackage{amsmath,amssymb,amsfonts}
\usepackage[linesnumbered,ruled,vlined]{algorithm2e}
\usepackage{graphicx}
\usepackage{multirow}
\usepackage{colortbl}
\usepackage{booktabs}
\usepackage{xcolor}
\usepackage{enumitem}
\usepackage{subfig}
\usepackage{tikz}
\usepackage{url}
\usepackage[font=small, skip=3pt]{caption}
\usepackage{pgfplots}
\usepackage{orcidlink}
\pgfplotsset{compat=1.18}
\usetikzlibrary{arrows.meta,shapes.geometric,positioning,calc,backgrounds,fit,decorations.pathreplacing}

\def\BibTeX{{\rm B\kern-.05em{\sc i\kern-.025em b}\kern-.08em
    T\kern-.1667em\lower.7ex\hbox{E}\kern-.125emX}}
\begin{document}

\title{EFaaS: A Quantum-Classical Serverless Entangled Scheduler for Hybrid Variational Algorithms}

\author{
Abolfazl Younesi\IEEEauthorrefmark{1}\orcidlink{0009-0003-0052-6475},
Nouhaila Innan\IEEEauthorrefmark{2}\IEEEauthorrefmark{3}\orcidlink{0000-0002-1014-3457},
Alberto Marchisio\IEEEauthorrefmark{2}\IEEEauthorrefmark{3}\orcidlink{0000-0002-0689-4776},
Muhammad Shafique\IEEEauthorrefmark{2}\IEEEauthorrefmark{3}\orcidlink{0000-0002-2607-8135}
\thanks{

\IEEEauthorrefmark{1}A. Younesi is Independent Researcher, Innsbruck, Austria. E-mail: Abolfazl.Younesi@ieee.org

\IEEEauthorrefmark{2}N. Innan, A. Marchisio, and M. Shafique are with the eBRAIN Lab, Division of Engineering, New York University Abu Dhabi, PO Box 129188, Abu Dhabi, UAE.

\IEEEauthorrefmark{3}N. Innan, A. Marchisio, and M. Shafique are also with the Center for Quantum and Topological Systems, NYUAD Research Institute, New York University Abu Dhabi, UAE. E-mails: \{nouhaila.innan, alberto.marchisio, muhammad.shafique\}@nyu.edu
}}


\maketitle
\begin{abstract}
As quantum computing enters the Utility Era, realizing near-term advantage relies heavily on Hybrid Variational Quantum Algorithms (VQAs). These algorithms require a tightly coupled, iterative loop between a classical CPU optimizer and a Quantum Processing Unit (QPU). However, current quantum cloud access models are bottlenecked by decoupled batch-queues that sever this loop, introducing massive Time-to-Next-Shot (TTNS) latency. This delay inflates convergence time from minutes to hours and exposes the computation to quantum hardware drift, degrading algorithmic fidelity. Unlike prior work that relies on resource-wasting static hardware reservations or state-oblivious, stateless functions, we propose \textit{EFaaS}, a novel serverless middleware designed specifically for hybrid quantum workflows. EFaaS fundamentally departs from existing architectures by treating classical parameter optimization and quantum circuit execution as entangled, session-aware events. Our main technical innovations are threefold: (1) a Calibration-Aware placement strategy that dynamically routes circuits to QPUs with warm calibration caches, circumventing cold-start penalties, (2) a Dual-Resource Fair Queuing scheduler that maximizes quantum utilization by strictly prioritizing active iterative loops, and (3) the ``\textit{EF-QuantumFuture}'' programming abstraction, a novel primitive enabling classical speculative execution to mask compute latency. 
Across the evaluated baselines, EFaaS achieves TTNS reductions of 11.4\%-94.3\%, QDC gains of 2.02\%-15.78\% points, and convergence speedups of 83.2\%-98.3\%, while eliminating drift penalties. 
\end{abstract}


\section{Introduction}
\IEEEPARstart{Q}{uantum}  computing has crossed the threshold into the ``Utility Era," a phase where quantum hardware can execute complex circuits beyond the reach of brute-force classical simulation \cite{kim2023evidence,10967435}. Realizing near-term quantum advantage in this era is heavily based on hybrid variational quantum algorithms (VQAs), such as the Variational Quantum Eigensolver (VQE) and the quantum approximation optimization algorithm (QAOA) \cite{cerezo_vqa_2021,10967435}. These algorithms operate fundamentally as a tightly coupled, iterative loop: a classical CPU optimizes parameterized variables, which are then evaluated as a cost function on a Quantum Processing Unit (QPU) through repeated circuit executions, or shots.

Despite algorithmic advances, such as the development of noise-resilient classical optimizers and hardware-efficient ansatz designs \cite{ramadhan2026hardware,gentini2020noise}, the systems-level architecture that supports these hybrid loops remains a significant bottleneck. Current quantum cloud access models operate primarily on batch-queue systems \cite{qiskit_runtime_2022}. In this paradigm, the iterative loop is repeatedly broken: a classical job submits a quantum circuit, waits in a queue, receives a result, classically optimizes, and resubmits. When queue times span seconds or minutes, a VQA requiring thousands of iterations can take days to converge.
Furthermore, this queueing latency introduces a critical physical vulnerability: quantum drift \cite{drift_qce_2020}. As the time between iterations extends (``cold starts"), QPU calibration data becomes stale. This drift degrades measurement fidelity, destroys the coherence of the optimization landscape, and forces costly system recalibrations \cite{proctor2020detecting}. Conversely, statically reserving scarce QPUs for the entire duration of a job while waiting for abundant CPUs to perform optimization is economically and computationally wasteful. Pure Functions-as-a-Service (FaaS) models \cite{nguyen_qfaas_2024} address classical scaling but fail to accommodate the stateful, synchronous demands of quantum hardware.

\textbf{Proposed approach.} We propose \textit{EFaaS}, a novel serverless scheduling middleware that sits between the quantum control plane and classical cloud orchestrators. EFaaS eliminates the traditional queuing bottleneck by treating QPU shots and CPU functions as entangled, session-aware events rather than isolated batch jobs. By implementing ``Session Awareness,'' the scheduler caches QPU calibration data while the classical CPU optimizes parameters. Furthermore, to mask classical computation latency, we introduce the ``\textit{EF-QuantumFuture}'' abstraction, a programming primitive that enables asynchronous variation and allows classical code to speculatively continue execution while awaiting QPU results.

\textbf{Contributions.} Our specific contributions are as follows:
\begin{itemize}[wide]
    \item We propose \textit{EFaaS} (Entangled Function-as-a-Service), a hybrid serverless execution model that co-schedules quantum circuit execution and classical parameter optimization as session-coupled events, drastically reducing the latency of the iterative loop (Time-to-Next-Shot).
    
    \item We design a \textit{Calibration-Aware Placement} strategy that routes circuits to QPUs with valid, iteration-specific calibration data, preventing cold starts and mitigating the impacts of quantum drift. The mechanism inherits its routing intuition from sticky-session affinity and locality-aware cluster schedulers, but reformulates the affinity validity predicate from a software invariant (cache hit, container warm) to a physical coherence-time threshold derived from qubit drift dynamics~\cite{proctor2020detecting}.
    
    \item We derive a \textit{Dual-Resource Fair Queuing} (DRFQ) scheduling policy that maximizes the \textit{Quantum Duty Cycle} (active QPU time) while preserving fair resource distribution for classical workloads. DRFQ extends Dominant Resource Fairness~\cite{ghodsi2011drf} to a setting where a single resource (QPU shots) is subject to a physical staleness constraint.
    
    \item We introduce \textit{EF-QuantumFuture}, a speculative-execution primitive for hybrid quantum-classical workflows. While quantum SDKs already expose asynchronous job-handle abstractions for deferred result retrieval, EF-QuantumFuture is, to our knowledge, the first to combine non-blocking dispatch with the overlap of strictly result-independent classical tasks, such as circuit transpilation and node warm-up, with pending QPU execution in a session-aware hybrid scheduler.
\end{itemize}

\textbf{Paper structure.} The remainder of this paper is organized as follows. Section \ref{sec:background} provides background. Section \ref{sec:related} reviews related work. Section \ref{sec:architecture} introduces the EFaaS system architecture. Section \ref{sec:scheduling} formulates the co-scheduling algorithms and placement strategies. Section \ref{sec:evaluation} presents performance evaluation, and Section \ref{sec:conclusion} concludes the paper.

\section{Background and Motivation}
\label{sec:background}

As quantum computing transitions from proof-of-concept experiments to the Utility Era, the focus is shifting toward algorithms that can run on near-term hardware without full fault tolerance. Achieving quantum advantage in this regime fundamentally depends on resolving the systems-level friction between classical and quantum resources.

\subsection{Hybrid Quantum-Classical Workflows}
\begin{figure}[h]
    \centering
    \includegraphics[width=1\linewidth]{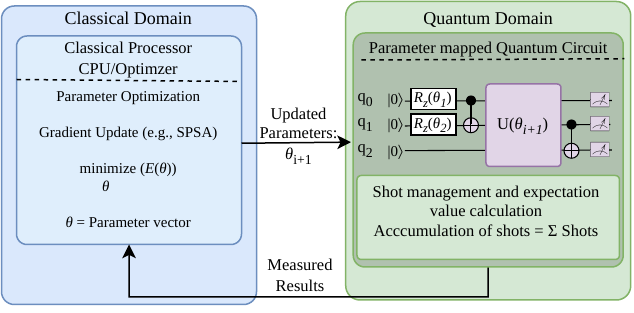}
    \caption{Iterative hybrid quantum-classical workflow for variational algorithms (VQE/QAOA), where a classical optimizer updates circuit parameters based on measurements from a quantum processor, forming a feedback loop between classical and quantum domains}
    \label{fig:HQCW}
\end{figure}

As shown in Fig.~\ref{fig:HQCW}, Variational Quantum Algorithms (VQAs) \cite{bharti2022noisy}, such as the Variational Quantum Eigensolver (VQE) \cite{mcclean2016theory} and the Quantum Approximate Optimization Algorithm (QAOA) \cite{choi2019tutorial}, are the primary candidates for near-term quantum utility \cite{kim2023evidence}. These algorithms do not run exclusively on a Quantum Processing Unit (QPU). Instead, they function as a tightly coupled, iterative loop between a classical processor and a quantum processor.
In a typical VQE workflow, a classical optimizer (e.g., SPSA \cite{spall2002multivariate}, COBYLA \cite{powell1994direct}) initializes a set of parameterized variables $\vec{\theta}$. These parameters define a quantum circuit (the ansatz), which is sent to the QPU. The QPU executes the circuit multiple times (shots) to measure the energy expectation value, representing the cost function of the problem landscape. {This result is returned to the classical CPU, which uses it to update the parameters to $\vec{\theta}_{i+1}$ via a stochastic finite-difference gradient approximation (e.g., SPSA, formed from two perturbed circuit evaluations per step) or a derivative-free direct-search update (e.g., COBYLA), depending on the chosen optimizer.} This loop must execute hundreds or thousands of times to achieve algorithmic convergence. Consequently, the performance of a VQA is dictated not just by quantum gate speeds, but by the round-trip communication and scheduling architecture binding the two domains.
\subsection{The Latency Bottleneck and TTNS}
The critical metric governing hybrid algorithm performance is the Time-to-Next-Shot (TTNS), the total latency incurred from the moment the classical CPU outputs new parameters to the moment the QPU begins evaluating the corresponding circuit. 
In the current quantum cloud paradigm, access is mediated by multi-tenant, batch-queue systems. When a VQA iterates, the updated circuit is submitted over the network and placed at the back of a FIFO (First-In-First-Out) or fair-share queue. The TTNS is thus defined as:
\begin{equation}
    TTNS = t_{net} + t_{queue} + t_{calib} + t_{qpu}
\end{equation}
where $t_{net}$ is network transmission, $t_{queue}$ is the wait time in the cloud provider's queue, $t_{calib}$ is potential hardware preparation delays, and $t_{qpu}$ is the quantum execution time.

Because $t_{queue}$ is highly variable and depends on cross-tenant traffic, it frequently dominates the TTNS equation, spanning from several seconds to minutes. For an algorithm requiring 5,000 iterations, a mere 30-second queue delay per iteration inflates the total runtime from a few minutes to over 40 hours. This latency severs the tight coupling required by the algorithm, transforming an iterative loop into a fragmented sequence of isolated batch jobs.

\subsection{Quantum Drift and Cold Starts}
The severe latency introduced by batch-queueing is not merely a performance issue; it is a profound physical vulnerability. Current superconducting and ion-trap QPUs are inherently noisy and highly sensitive to environmental fluctuations. Consequently, the physical properties of the qubits, such as $T_1$ relaxation times, $T_2$ coherence times, and gate error rates, drift continuously. 

To counteract this, cloud providers periodically run calibration sequences to update the control microwave pulses ($C_q$). However, this calibration data is valid only within a limited temporal window, denoted by $\tau_{\text{drift}}$. If the delay between VQA iterations ($t_{\text{queue}}$) exceeds $\tau_{\text{drift}}$, the QPU undergoes a ``cold start.'' Cold starts force the system to either pause the workload to undergo a costly recalibration cycle ($t_{\text{calib}}$) or, worse, execute the circuit using stale calibration data. If the circuit runs on stale data, the measurement fidelity drops and the cost function $\langle H \rangle_t$ becomes \emph{non-stationary} across iterations: identical parameter vectors $\vec{\theta}$ return different expectation values at different wall-clock times. This non-stationarity violates the i.i.d.\ sampling assumption underlying stochastic optimizers such as SPSA~\cite{
spall2002multivariate} and desynchronizes finite-difference gradient estimates, forcing the classical node to execute more iterations to converge, if it converges at all~\cite{proctor2020detecting}. We emphasize that this drift-induced non-stationarity is distinct from the barren-plateau phenomenon, which is a static property of the parameterized ansatz under random initialization, and from noise-induced barren plateaus, which arise from coherent or incoherent gate-noise channels rather than inter-iteration calibration aging.

\subsection{Resource Asymmetry and Allocation Inefficiencies}
Solving the latency and drift bottlenecks requires reconciling a massive asymmetry in resource availability. Classical compute resources (CPUs, FaaS nodes) are practically infinite and highly elastic. In contrast, state-of-the-art QPUs are scarce, expensive, and strictly localized.
Current architectures attempt to bypass queueing latency via \textit{Static Reservation}, in which a user pays to exclusively lock the physical QPU for the duration of the VQA job. While this drives $t_{queue} \to 0$, it is economically unscalable for public clouds and computationally highly inefficient. During a static reservation, whenever the classical CPU is performing complex parameter updates ($t_{cpu}$), the locked QPU remains idle. 
Conversely, relying solely on pure stateless Functions-as-a-Service (FaaS) \cite{nguyen_qfaas_2024} for classical compute fails to address the stateful requirements of quantum hardware. Stateless classical architectures inherently drop the QPU session context between invocations, defaulting back to cold starts and drift penalties. Therefore, a novel architectural paradigm is required, one that preserves the stateful session context of scarce quantum hardware while dynamically co-scheduling abundant classical functions to minimize idle time and maximize the Quantum Duty Cycle.

\section{Related Work}
\label{sec:related}

EFaaS sits at the intersection of four active research areas:
hybrid quantum-classical computing frameworks, quantum cloud resource
management, classical serverless scheduling, and variational algorithm
optimization. We survey each in turn and position our work against the
state of the art. Table~\ref{tab:related_comparison} provides a
structured comparison across the key dimensions of our problem
formulation.

\subsection{Hybrid Quantum-Classical Frameworks}
\label{subsec:related_hybrid}

The VQE~\cite{peruzzo2014vqe} and the QAOA~\cite{farhi2014qaoa} established the hybrid quantum-classical loop as the dominant execution paradigm for near-term quantum devices. Both algorithms depend critically on tight CPU-QPU coupling: the classical optimizer must receive QPU results and resubmit updated circuits with minimal latency. Subsequent work refined the algorithmic side gradient estimation via parameter-shift rules~\cite{mitarai2018quantum}, noise-aware ansatz design~\cite{kandala2017hardware}, and adaptive circuit construction~\cite{grimsley2019adaptive} but uniformly assumed idealized, low-latency QPU access. None of these works model or mitigate the scheduling latency introduced by real cloud quantum access models, which is the central gap EFaaS addresses.

\begin{table*}[t]
\centering
\setlength{\extrarowheight}{0pt}
\setlength{\aboverulesep}{0pt}
\setlength{\belowrulesep}{0pt}
\caption{Comparison of hybrid quantum-classical execution approaches across key scheduling dimensions. \textbf{Yes} = fully supported; \textit{Partial} = partially supported; No = not supported.}
\label{tab:related_comparison}
\resizebox{\linewidth}{!}{%
\begin{tabular}{lccccccc} 
\toprule
\rowcolor[rgb]{0.749,0.749,0.749} \textbf{Approach} & \textbf{\shortstack{Session-Aware /\\Calib. Caching}} & \textbf{\shortstack{Calibration-Aware\\Placement}} & \textbf{\shortstack{Dual-Resource\\Co-Scheduling}} & \textbf{\shortstack{Speculative /\\Latency Masking}} & \textbf{\shortstack{Serverless /\\FaaS-like}} & \textbf{\shortstack{Formal Fairness\\Guarantee}} & \textbf{\shortstack{Low TTNS for\\Active VQA Loops}} \\ 
\midrule
VQE / QAOA~\cite{peruzzo2014vqe,farhi2014qaoa} & No & No & No & No & No & No & No \\
Qiskit Runtime / Serverless~\cite{qiskit_serverless_2024} & \textit{Partial} (sessions) & No & \textit{Partial} (priority grouping) & No & Yes (recent) & No & Moderate \\
Amazon Braket Hybrid Jobs~\cite{braket_hybrid_2021} & \textit{Partial} (job lifetime) & No & \textit{Partial} (priority during job) & No & Container-based & No & Moderate \\
Pilot-Quantum~\cite{mantha_pilot_quantum_2024} & No & No & Yes (multi-level) & No & No & No & Low-Moderate \\
QFaaS / Traditional FaaS~\cite{nguyen_qfaas_2024,jonas2019cloud} & No & No & No & No & Yes & No & Low \\
Other Hybrid Middleware~\cite{faro_middleware_quantum_2023,weder_analysis_rewrite_2022} & No & \textit{Partial} & \textit{Partial} & No & No & No & Low \\
Noise-Aware Placement~\cite{murali2019noise} & No & \textit{Partial} (compile-time) & No & No & No & No & No \\
Quantum-Cloud Workloads~\cite{ravi2021quantum} & No & No & No & No & No & No & No \\
FaaS Cold-Start Mitigation~\cite{shahrad2020serverless,gunasekaran2020fifer} & \textit{Partial} (keep-alive) & No & No & No & Yes & No & No \\
Quantum Pipelining~\cite{shi2020resource} & No & No & No & \textit{Partial} (compile) & No & No & No \\
Static Reservation & Yes (exclusive) & \textit{Partial} & No & No & No & No & High (wasteful) \\ 
\midrule
\rowcolor[rgb]{0.678,0.839,0.678} \textbf{EFaaS (Proposed)} & \textbf{Yes} & \textbf{Yes} & \textbf{Yes (DRFQ)} & \textbf{Yes (\textit{EF-QuantumFuture})} & \textbf{Yes} & \textbf{Yes} & \textbf{High} \\
\bottomrule
\end{tabular}
}
\end{table*}
On the platform side, IBM's Qiskit Runtime~\cite{qiskit_runtime_2022} introduced the \texttt{Session} primitive to reduce inter-job queuing overhead by batching circuit submissions within a reserved QPU window. {Per current IBM documentation, a Session grants the submitting workload \emph{dedicated and exclusive} QPU access for a configurable time-to-live once the first job in the session begins execution, with subsequent jobs in the session prioritized by the provider's internal scheduler. This reduces inter-job queuing overhead relative to independent batch submissions, but exclusivity is held for the full session duration rather than released between individual shots, so the QPU remains reserved (and billed) during classical-optimization intervals structurally closer to the tradeoff of our Static Reservation (SR) baseline than to cooperative, TTL-bounded leasing.} Qiskit Serverless~\cite{qiskit_serverless_2024} (2024) extends this further with a programming model for long-running hybrid workloads across distributed CPU, GPU, and QPU resources. {Qiskit's primitive jobs (\texttt{Sampler}/\texttt{Estimator}) are asynchronous by default, so users already have the raw capability to structure classical code around a pending result. This is an SDK-level dispatch primitive, however, not a scheduling policy. It does not itself define which classical work is safe to overlap, does not track physical calibration validity against a drift threshold, and does not arbitrate fairness across competing sessions under contention.} While Sessions reduces cold-start penalties within a single job and Qiskit Serverless improves scalability, both remain decoupled at the scheduler level: classical optimizer invocations and QPU shots are not treated as jointly scheduled resources, calibration state is not explicitly cached with TTL-aware eviction, and variable Time-to-Next-Shot (TTNS) persists under contention, leaving active loops exposed to calibration drift. {EFaaS's contribution relative to these production systems is therefore the co-scheduling \emph{policy} layered on top of capabilities that existing SDKs already expose: calibration-TTL-indexed placement, a dual-resource fair-queuing formulation with an explicit fairness weighting, and a scoped, result-independent classical-overlap primitive. We caveat that provider-internal scheduler behavior beyond what is documented is not observable to us, so this comparison is necessarily based on published documentation rather than internal implementation details.} Amazon Braket Hybrid Jobs~\cite{braket_hybrid_2021} similarly enables hybrid orchestration by running user code in managed containers with priority QPU access throughout a job's lifetime, reducing queue times for iterative tasks. However, it lacks calibration-aware routing to warm-cached backends and provides no mechanism for speculative classical execution during QPU wait periods.

Pilot-Quantum~\cite{mantha_pilot_quantum_2024} provides a unified middleware abstraction for resource, workload, and task management across CPUs, GPUs, and QPUs, supporting variational algorithms via task parallelism and integration with Qiskit and PennyLane. It emphasizes modular orchestration and adaptive allocation, but places limited emphasis on drift-specific placement or low-latency session continuity. Other hybrid middleware proposals address orchestration patterns~\cite{faro_middleware_quantum_2023}, workflow rewriting~\cite{weder_analysis_rewrite_2022}, and circuit-level concurrency for accelerating variational workloads~\cite{resch_accelerating_vqa_2021}, yet none jointly optimizes calibration validity, dual-resource fairness, and speculative execution within a unified scheduler.

\subsection{Quantum Cloud Resource Management}
\label{subsec:related_cloud}

Murali et al.~\cite{murali2019noise} demonstrated that QPU backend selection significantly affects circuit fidelity, motivating noise-aware compilation and placement at the compilation layer. Their work does not address the runtime scheduling of iterative jobs or the temporal evolution of calibration validity between submissions. Paler et al.~\cite{paler2021machine} proposed machine-learning approaches for quantum circuit layout optimization, focusing on reducing compiled circuit depth. Ravi et al.~\cite{ravi2021quantum} analyzed quantum-cloud job and machine characteristics, including queueing delays, execution times, and machine utilization, without coupling these observations to classical optimizer state or modeling inter-iteration drift accumulation.

Static QPU reservation avoids queuing but wastes QPU time during classical optimization phases and locks out concurrent users. Pure batch models incur high TTNS and fidelity loss from drift~\cite{drift_qce_2020}. EFaaS occupies the space between these extremes through session leasing, calibration caching, and dual-resource co-scheduling.

\subsection{Serverless and FaaS Scheduling}
\label{subsec:related_faas}

Serverless computing and FaaS platforms such as AWS Lambda~\cite{jonas2019cloud} and Knative~\cite{knative2023} eliminate infrastructure management by dynamically provisioning short-lived function containers. The cold-start latency inherent to FaaS has been extensively studied~\cite{wang2018peeking,shahrad2020serverless}, with mitigations including keep-alive policies~\cite{gunasekaran2020fifer}, predictive pre-warming~\cite{lin2019mitigating}, and checkpoint-restore techniques~\cite{du2020catalyzer}. These classical FaaS insights directly inform our Session Awareness Engine: the TTL-based calibration cache is structurally analogous to a keep-alive policy, adapted for the physical decay dynamics of qubit coherence rather than container memory state.
Emerging quantum-serverless proposals, such as QFaaS~\cite{nguyen_qfaas_2024}, explore FaaS abstractions for quantum circuits and address classical scaling effectively, but struggle with stateful quantum synchronization and the scarcity of QPU resources. Critically, no existing FaaS or quantum-FaaS framework is aware of the iterative, coupled nature of hybrid workloads, defines QPU shot budgets as a schedulable resource share, or provides a speculative execution primitive for the classical optimizer. EFaaS introduces all three of these missing dimensions into the serverless execution model. Dominant Resource Fairness (DRF)~\cite{ghodsi2011drf} provides the theoretical foundation for our DRFQ scheduler. DRF achieves max-min fairness in multi-resource settings. Our contribution extends DRF to a domain where one resource (QPU shots) carries a physical staleness constraint, requiring the joint optimization of fairness and calibration validity, a problem not considered in any prior DRF or quantum scheduling literature.

\subsection{Speculative and Asynchronous Execution}
\label{subsec:related_speculative}

Speculative execution is well-established in classical processor microarchitecture ~\cite{hennessy2011computer} and distributed systems~\cite{dean2013tail}, where tasks execute ahead of confirmed dependencies to hide latency. In quantum computing, Shi et al.~\cite{shi2020resource} explored pipelining a quantum circuit compilation stage, and Ding et al.~\cite{ding2020systematic} proposed frequency-aware compilation for crosstalk mitigation, but neither extends speculative execution to the optimizer parameter-update loop at runtime. {The \texttt{QuantumFuture} abstraction introduced in EFaaS targets this gap by overlapping strictly result-independent classical work (e.g., next-circuit transpilation, classical-node warm-up) with QPU execution time, rather than by speculating on the optimizer's parameter update itself, which necessarily depends on the pending measurement $\langle H\rangle_i$.} This addresses the latency-masking gap identified in the prior hybrid frameworks surveyed above.

\subsection{Comparative Summary}
\label{subsec:related_comparison}


Table~\ref{tab:related_comparison} summarizes how EFaaS differs from the most closely related systems across seven key dimensions. No prior system simultaneously addresses all seven: session-aware calibration caching, calibration-aware placement, dual-resource co-scheduling, speculative classical execution, serverless deployment, formal fairness guarantees, and low TTNS for
active VQA loops.
EFaaS advances beyond all prior work by fully coupling classical and quantum scheduling events through four synergistic mechanisms: (i) TTL-aware session caching in the SAE eliminates cold-start recalibration, (ii) calibration-aware placement routes circuits to backends with valid cached qubit fidelity data, (iii) DRFQ enforces max-min fairness jointly over QPU shots and CPU cycles, and (iv) the \texttt{QuantumFuture} abstraction overlaps speculative classical optimization with QPU execution, achieving near-zero TTNS for active VQA loops while preserving serverless
elasticity and drift resilience.

\section{EFaaS Architecture}
\label{sec:architecture}

To resolve the deep latency and coherence bottlenecks inherent in current hybrid quantum-classical workflows, we introduce \textit{EFaaS}. This serverless architecture fundamentally redefines the relationship between classical parameter optimization and quantum circuit execution, transitioning from a decoupled batch-queue model to a tightly bound, synchronous event-driven paradigm.

\subsection{System Overview}
The EFaaS system can act as an intelligent middleware layer situated between a classical cloud orchestrator and the Quantum Control Plane of the QPU provider. 

\begin{figure}[htbp]
    \centering
    \includegraphics[width=1\linewidth]{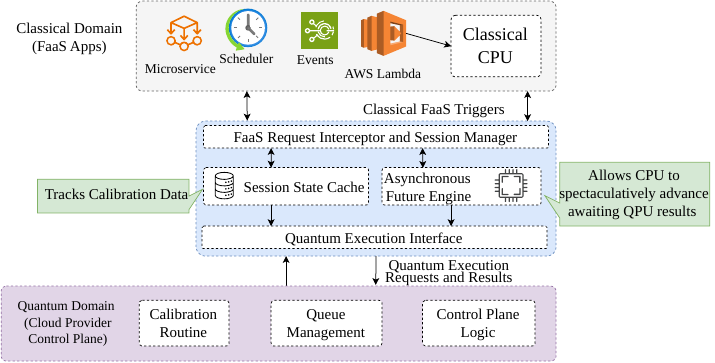}
    \caption{EFaaS System Architecture. The middleware intercepts Classical FaaS triggers and Quantum Execution requests (Control Plane). The ``Session State Cache" maintains QPU calibration data ($C_q$) while the ``Asynchronous Future Engine" allows the CPU to speculatively advance while awaiting QPU results.}
    \label{fig:architecture}
\end{figure}

Traditional architectures treat the classical CPU optimizer and the QPU evaluating the cost function as separate, asynchronous entities. EFaaS unifies them using a co-scheduling controller. When a hybrid algorithm (such as VQE) is submitted, the orchestrator allocates a classical function instance. Simultaneously, the EFaaS middleware negotiates a high-priority, session-aware window with the Quantum Control Plane, creating an ``entangled" execution loop.

\subsection{Mathematical Formulation and Notation}

To rigorously define the latency advantages of EFaaS, we formulate the Time-to-Next-Shot (TTNS) under both standard and proposed architectures. The key system variables are defined in Table \ref{tab:variables}.

\begin{table}[htbp]
\caption{System Variables and Notation}
\label{tab:variables}
\centering
\begin{tabular}{ll}
\hline
\textbf{Symbol} & \textbf{Definition} \\ \hline
$L_{std}$ & Latency per iteration (Standard) \\
$L_{ent}$ & Latency per iteration (EFaaS) \\
$t_{cpu}$ & Execution time of classical optimization \\
$t_{qpu}$ & QPU active execution time (shots) \\
$t_{queue}$ & Time spent in standard cloud queue \\
$t_{calib}$ & Time required to re-calibrate QPU \\
$\tau_{drift}$ & Quantum coherence/drift time threshold \\
$C_q$ & Calibration data matrix for target qubits \\
$E_s$ & Session awareness active flag $\{0,1\}$ \\
$t_{async}$ & Speculative classical execution time \\ \hline
\end{tabular}
\end{table}

In a standard cloud access model, the total latency for a single iteration of a VQA is the linear sum of classical computation, queueing delay, potential recalibration, and quantum execution:
\begin{equation}
    L_{std} = t_{cpu} + t_{queue} + \delta \cdot t_{calib} + t_{qpu}
\end{equation}
where $\delta = 1$ if $t_{queue} > \tau_{drift}$ (indicating a cold start due to stale calibration data), and $0$ otherwise. Because $t_{queue}$ is often highly variable and frequently exceeds $\tau_{drift}$ in public quantum clouds, algorithms are heavily penalized by $t_{calib}$.

\subsection{Session Awareness and Calibration Caching}
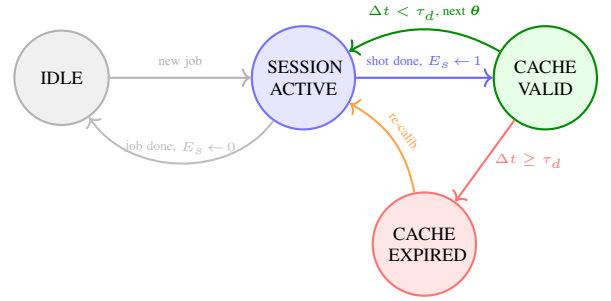
\begin{figure}[htbp]
\centering
\begin{tikzpicture}[font=\small,
  every node/.style={},
  st/.style={circle,draw=gray!60,fill=gray!12,thick,minimum size=1.05cm,
             text width=0.95cm,align=center,font=\scriptsize},
  stB/.style={st,draw=blue!55,fill=blue!10},
  stG/.style={st,draw=green!55!black,fill=green!10},
  stR/.style={st,draw=red!55,fill=red!10},
]
  \node[st]  (idle)    at (0,0)      {IDLE};
  \node[stB] (active)  at (3.2,0)    {SESSION\\ACTIVE};
  \node[stG] (cached)  at (6.4,0)    {CACHE\\VALID};
  \node[stR] (expired) at (4.8,-2.2) {CACHE\\EXPIRED};

  \draw[->,thick,gray!60]   (idle) -- (active)
    node[midway,above,font=\tiny]{new job};
  \draw[->,thick,blue!60]   (active) -- (cached)
    node[midway,above,font=\tiny]{shot done, $E_s\!\leftarrow\!1$};
  \draw[->,thick,red!55]    (cached) -- (expired)
    node[midway,right,font=\tiny]{$\Delta t \geq \tau_d$};

  \draw[->,thick,green!55!black]
    (cached) to[bend right=30]
    node[midway,sloped,above,font=\tiny]{$\Delta t < \tau_d$,\;next $\boldsymbol{\theta}$}
    (active);

  \draw[->,thick,orange!70]
    (expired) to[bend right=25]
    node[midway,sloped,above,font=\tiny]{re-calib.}
    (active);

  \draw[->,thick,gray!50]
    (active) to[bend left=55]
    node[midway,sloped,above,font=\tiny]{job done, $E_s\!\leftarrow\!0$}
    (idle);
\end{tikzpicture}
\caption{Session Awareness Engine state machine. A Hot Iterator cycles between SESSION ACTIVE and CACHE VALID provided each inter-shot delay $\Delta t$ remains within $\tau_d = \tau_\text{drift}$. If the cache expires, the engine triggers a localized recalibration before resuming the active session.}
\label{fig:session_fsm}
\end{figure}

The core mechanism of EFaaS is \textit{Session Awareness} (see Fig.\ref{fig:session_fsm}). Instead of tearing down the quantum environment after $t_{qpu}$ is completed, the middleware asserts the session flag $E_s = 1$. This instructs the Quantum Control Plane to hold the calibration data matrix $C_q$ in a localized cache. 

The scheduler dynamically calculates the maximum allowable classical compute window before drift forces a recalibration:
\begin{equation}
    t_{cpu} \leq \tau_{drift} - \epsilon
\end{equation}
where $\epsilon$ is a system safety margin. As long as the classical CPU returns the next parameterized circuit within $\tau_{drift}$, the QPU executes the shots immediately. By elevating the priority of circuits returning to a warm cache, EFaaS drives $t_{queue} \to 0$ and $\delta \to 0$.

\subsection{The ``\textit{EF-QuantumFuture}" Abstraction}
\begin{figure}[htbp]
\centering
\begin{tikzpicture}[font=\small, x=1cm, y=1cm]
  \def\xC{0.0} \def\xM{3.0} \def\xQ{6.0}

  \foreach \xp/\lbl in {\xC/{Classical\\Optimizer}, \xM/{EFaaS\\Middleware}, \xQ/{QPU}}{
    \node[draw,fill=gray!15,rounded corners,font=\tiny,minimum width=1.75cm,
          minimum height=0.6cm,align=center] at (\xp,0.35){\lbl};
    \draw[gray!45,dashed,thin](\xp,0.0)--(\xp,-6.0);
  }
  
  \fill[blue!12](\xC-0.10,0.0)rectangle(\xC+0.10,-6.0);
  \draw[blue!35,thin](\xC-0.10,0.0)rectangle(\xC+0.10,-6.0);
  \fill[gray!18](\xM-0.10,-0.7)rectangle(\xM+0.10,-5.1);
  \draw[gray!40,thin](\xM-0.10,-0.7)rectangle(\xM+0.10,-5.1);
  \fill[green!12](\xQ-0.10,-1.6)rectangle(\xQ+0.10,-3.8);
  \draw[green!45,thin](\xQ-0.10,-1.6)rectangle(\xQ+0.10,-3.8);
  
  \draw[->,thick](\xC+0.10,-0.9)--(\xM-0.10,-0.9);
  \node[above,font=\tiny]at(1.5,-0.9){submit $\boldsymbol{\theta}_i$, circuit};
  \draw[->,thick](\xM+0.10,-1.6)--(\xQ-0.10,-1.6);
  \node[above,font=\tiny]at(4.5,-1.6){dispatch};
  \draw[->,dashed,thick](\xM-0.10,-2.1)--(\xC+0.10,-2.1);
  \node[above,font=\tiny]at(1.5,-2.1){\texttt{Future f} (async)};

  \fill[violet!10](\xC-0.9,-2.35)rectangle(\xC-0.12,-3.75);
  \draw[violet!40,dashed](\xC-0.9,-2.35)rectangle(\xC-0.12,-3.75);
  \node[font=\tiny,violet!65,align=center]at(\xC-0.51,-3.05){specul.\\work\\$t_\text{async}$};

  \node[font=\tiny,green!60!black,rotate=90]at(\xQ,-2.7){QPU ($t_\text{qpu}$)};

  \draw[->,thick](\xQ-0.10,-3.8)--(\xM+0.10,-3.8);
  \node[above,font=\tiny]at(4.5,-3.8){result $\langle H\rangle$};
  \draw[->,thick](\xM-0.10,-4.4)--(\xC+0.10,-4.4);
  \node[above,font=\tiny]at(1.5,-4.4){\texttt{f.resolve($\langle H\rangle$)}};
  \node[font=\tiny,blue!65,anchor=west]at(\xC+0.13,-4.95){$\boldsymbol{\theta}_{i+1}\!\leftarrow\!\text{step}()$};
  \draw[->,thick](\xC+0.10,-5.55)--(\xM-0.10,-5.55);
  \node[above,font=\tiny]at(1.5,-5.55){submit $\boldsymbol{\theta}_{i+1}$};
  \draw[->,gray!55,thin](-1.25,0.0)--(-1.25,-6.0);
  \node[font=\tiny,gray,rotate=90]at(-1.5,-3.0){time $\longrightarrow$};
\end{tikzpicture}
\caption{\textit{EF-QuantumFuture} execution sequence. The Middleware returns an asynchronous \texttt{Future f} immediately upon submission, allowing the classical optimizer to perform speculative work ($t_\text{async}$) concurrently with QPU execution ($t_\text{qpu}$). The \texttt{f.resolve()} call delivers $\langle H\rangle$ once the QPU completes, and the optimizer proceeds to $\boldsymbol{\theta}_{i+1}$ without any blocking wait.}
\label{fig:qfuture_seq}
\end{figure}
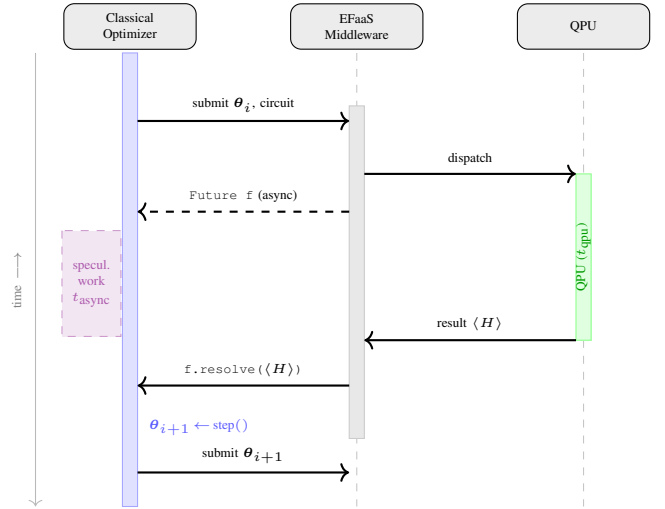

To further minimize the perceived latency and decouple the strict synchronous waiting of the classical node, we introduce the \textit{\textit{EF-QuantumFuture}} programming primitive (see Fig. \ref{fig:qfuture_seq}).

In traditional VQAs, the classical process blocks entirely while the QPU evaluates the energy expectation value $\langle H \rangle$. The \textit{EF-QuantumFuture} object returns an immediate asynchronous promise to the classical optimizer. {This allows the classical function to continue with \emph{result-independent} work such as precompiling or transpiling the next candidate circuit under the current calibration map, or warming up non-quantum-dependent classical nodes while $t_{qpu}$ runs. Because the next parameter update $\boldsymbol{\theta}_{i+1}$ is itself computed from the pending measurement $\langle H\rangle_i$, EF-QuantumFuture does not speculate on the optimizer step. It eliminates only the dispatch-adjacent latency that would otherwise serialize after $\langle H\rangle_i$ arrives.}

If $t_{async}$ represents the time spent executing speculative classical operations concurrently with the QPU, the iteration latency under EFaaS is reduced to:
\begin{equation}
    L_{ent} = \max(t_{cpu} - t_{async}, 0) + t_{qpu} + t_{net}
\end{equation}
where $t_{net}$ is the negligible network transmission overhead. By co-scheduling the resources and overlapping $t_{async}$ with $t_{qpu}$, EFaaS strictly bounds the execution time and completely isolates the iterative loop from external queueing disruptions.
{Because \texttt{f.resolve()} has a single deterministic outcome (delivery of $\langle H\rangle$) and no speculative optimizer state is created, no rollback or reconciliation mechanism is required; we therefore do not claim a speculation/commit/abort contract for the optimizer state.}

\section{Co-Scheduling and Algorithmic Framework}
\label{sec:scheduling}

The fundamental innovation of EFaaS is its departure from independent, isolated job queues for classical and quantum resources. To realize the latency reductions modeled in Section \ref{sec:architecture}, we introduce a co-scheduling framework that combines Dual-Resource Fair Queuing with a Calibration-Aware Placement Strategy.

\subsection{Calibration-Aware Placement Strategy}
\label{subsec:placement}
In standard architectures, a quantum circuit is routed to the first available QPU  matching its topology requirements. EFaaS introduces \textit{Calibration-Aware  Placement}, a sticky-session affinity policy specialized to the physical-decay regime of QPU calibration. Conceptually analogous to node-locality preferences in  YARN~\cite{vavilapalli2013yarn} and topology-aware routing in  Kubernetes~\cite{kubernetes2023topology}, our placement engine prefers the QPU  that holds the relevant warm state. Unlike classical affinity, however, the validity predicate is governed not by software invariants (cache presence,  container liveness) but by a physical coherence-time threshold $\tau_{\text{drift}}$  derived from qubit drift dynamics~\cite{proctor2020detecting}.

The placement engine actively tracks the calibration state $C_q$ of the physical hardware. If a hybrid algorithm is executing an iterative loop, the target QPU's quantum state drifts over time. We define a validity threshold $\tau_{\text{drift}}$ that specifies the maximum time window after which a QPU's calibration data is considered stale. The scheduler maintains a mapping of active sessions to specific QPUs. When a new circuit $q_i$ from an active iteration arrives, the placement engine attempts to route it to the exact QPU $u_k$ utilized in the previous iteration, provided that the elapsed time $\Delta t$ satisfies $\Delta t < \tau_{\text{drift}}$. This prevents the severe latency penalty of cold-start recalibrations.

\subsection{Dual-Resource Fair Queuing}
To prioritize active quantum loops without starving classical workloads or independent quantum batch jobs, we define the concept of a ``Hot Iterator." A Hot Iterator is an active VQA session where the classical CPU is currently computing the next set of parameters, and the QPU is holding its calibration state in cache ($E_s = 1$).

We define the scheduling objective to maximize the Quantum Duty Cycle ($QDC$), which is the ratio of active quantum execution time to total wall-clock time:
\begin{equation}
    \max QDC = \frac{1}{T} \sum_{i=1}^{N} t_{qpu}^{(i)}
\end{equation}

To achieve this, the scheduler calculates a dynamic priority score $\rho_i$ for each incoming job $i$:
\begin{equation}
    \rho_i = \alpha \cdot \mathbb{I}(E_s^{(i)} = 1) + \beta \cdot \left( \frac{\tau_{drift} - \Delta t_{wait}}{\tau_{drift}} \right) + \gamma \cdot W_i
\end{equation}
where $\mathbb{I}$ is the indicator function for an active session, $\Delta t_{wait}$ is the time elapsed since the last shot, $W_i$ is a standard fair-share weight to prevent starvation of batch jobs, and $\alpha, \beta, \gamma$ are tunable hyperparameters. Hot Iterators receive a massive boost via $\alpha$, ensuring that as soon as the classical node yields a parameter update, the QPU executes it immediately.

\subsection{Entangled Scheduling Algorithm}
The logic governing the EFaaS middleware is formalized in Algorithm \ref{alg:co_scheduler}.

\begin{algorithm}[hbt!]
\caption{Calibration-Aware Dual-Resource Scheduling}
\label{alg:co_scheduler}
\LinesNumbered
\SetAlgoNoEnd
\footnotesize
\KwIn{Job queue $J$, Available QPUs $U$, Current time $t_{now}$} \label{alg:in}
\KwOut{Assigned QPU $u^*$ or Classical Node $c^*$} \label{alg:out}
\While{$J$ is not empty}{ \label{alg:while}
    Pop highest priority job $j$ from $J$ based on $\rho$\; \label{alg:pop}
    \eIf{$j$ is Quantum Circuit $q$}{ \label{alg:if_q}
        \If{$q$ has active session flag $E_s == 1$}{ \label{alg:check_session}
            $u^* \leftarrow$ find\_cached\_qpu($q, U$)\; \label{alg:find_cache}
            \eIf{$(t_{now} - u^*.last\_calib) < \tau_{drift}$}{ \label{alg:check_drift}
                Assign $q$ to $u^*$ with Preemptive Priority\; \label{alg:assign_high}
                Update $u^*.last\_calib \leftarrow t_{now}$\; \label{alg:update_calib}
            }{ \label{alg:else_drift}
                Trigger localized recalibration on $u^*$\; \label{alg:recalib}
                Assign $q$ to $u^*$\; \label{alg:assign_after_recalib}
            } \label{alg:end_drift_check}
        } \label{alg:end_session_check}
        \Else{ \label{alg:else_session}
            $u^* \leftarrow$ standard\_fair\_queue($q, U$)\; \label{alg:assign_std}
        } \label{alg:end_else_session}
    }{ \label{alg:else_q}
        $c^* \leftarrow$ assign\_classical\_node($j$)\; \label{alg:assign_c}
        Initialize \textit{EF-QuantumFuture} async object for $c^*$\; \label{alg:notify}
    } \label{alg:end_if_q}
} \label{alg:endwhile}
\end{algorithm}

\textbf{Algorithm Explanation:} 
The algorithm continuously processes incoming hybrid workloads, taking inputs of jobs and available resources. The main event loop begins at Line \ref{alg:while}, popping jobs sorted by the priority score $\rho$ defined in Equation 6 (Line \ref{alg:pop}). 
If the job is identified as a quantum circuit (Line \ref{alg:if_q}), the scheduler checks if it is part of a Hot Iterator by verifying the session flag $E_s$ (Line \ref{alg:check_session}). For active sessions, the middleware attempts to locate the specific QPU that holds the cached calibration data (Line \ref{alg:find_cache}). Crucially, Lines \ref{alg:check_drift} through \ref{alg:update_calib} evaluate the quantum drift threshold. If the time elapsed is strictly less than $\tau_{drift}$, the circuit is preemptively assigned, bypassing standard queues and avoiding cold starts. If the drift threshold has been exceeded (Line \ref{alg:else_drift}), a localized, partial recalibration is triggered before execution (Lines \ref{alg:recalib}-\ref{alg:assign_after_recalib}).
Standard quantum batch jobs lacking an active session flag fall back to standard fair queuing (Lines \ref{alg:else_session}-\ref{alg:assign_std}). Conversely, if the popped job is a classical optimization task (Line \ref{alg:else_q}), it is assigned to a classical FaaS node (Line \ref{alg:assign_c}). Immediately upon this assignment, the \textit{\textit{EF-QuantumFuture}} asynchronous object is initialized (Line \ref{alg:notify}), allowing the classical node to begin speculative execution while the QPU resolves the subsequent circuit execution.

\section{Experimental Evaluation}
\label{sec:evaluation}

To validate the performance of the EFaaS architecture, we constructed a comprehensive simulation environment that models the execution of hybrid VQE algorithms across distributed classical and quantum infrastructure.

\textbf{Evaluation Methodology and Testbed.}
Evaluating a unified quantum-classical serverless model requires co-scheduling visibility that is difficult to obtain from production cloud control planes. We therefore use a discrete-event simulator implemented with SimPy and instrumented with Qiskit Aer-based quantum evaluations. The simulator explicitly models queueing, drift, calibration, pilot startup/overhead, and background traffic, and records per-iteration TTNS, convergence timing, QDC, and drift penalties. Queue delay is modeled as a high-variance lognormal process ($\mu=3.5,\sigma=0.8$), and background contention is modeled as a Poisson process ($\lambda=0.05$ jobs/s) with exponential QPU service times (mean 5 s). The concrete software environment and core runtime settings are listed in Table \ref{tab:testbed}.

\begin{table}
\centering
\caption{Experimental Testbed and Software Environment}
\label{tab:testbed}
\begin{tabular}{ll} 
\hline
\rowcolor[rgb]{0.753,0.753,0.753} \textbf{Component} & \textbf{Specification / Version} \\ 
\hline
Language Runtime & Python 3.11.x \\
Quantum SDK & Qiskit \\
Quantum Primitive & Qiskit Aer Estimator \\
Queueing Model & Lognormal ($\mu=3.5,\sigma=0.8$) \\
Background Traffic & Poisson arrivals ($\lambda=0.05$/s) \\
QPU Pool Size & $N_{QPU}=3$ \\
Classical Pool Size & $N_{classical}=4$ \\
Simulation Horizon & $T=3000$ simulated seconds \\
\hline
\end{tabular}

\end{table}

\textbf{Workload and Hyperparameters.}
Our workload is an iterative VQE-style benchmark driven by SPSA with up to 1000 iterations per run. For observables, we use a LiH Hamiltonian and a transverse-field Ising-chain Hamiltonian for larger circuits.

We benchmark EFaaS against four architectural baselines currently utilized in quantum-cloud ecosystems:
\begin{enumerate}[wide]
    \item \textbf{Standard Batch-Queue (SBQ):} The default model where each iteration is a discrete job in a FIFO queue.
    \item \textbf{Pure FaaS (PF):} A decoupled model where the optimizer runs statelessly, but quantum jobs enter a standard queue.
    \item \textbf{Static Reservation (SR):} A dedicated access model locking the QPU exclusively for the job's duration.
    \item \textbf{Pilot-Quantum (PQ) \cite{mantha_pilot_quantum_2024}:} A pilot-job model with one startup phase followed by low-overhead warm task dispatch.
\end{enumerate}

The algorithmic and scheduling hyperparameters governing both the VQE payload and the Dual-Resource scheduler are defined in Table \ref{tab:hyperparams}.
For sensitivity analysis, we sweep $\alpha\in\{0,10,50,100,200\}$, $\beta\in\{0,1,5,10,20\}$, $\gamma\in\{0.1,0.5,1.0,2.0,5.0\}$, and $\tau_{drift}\in\{60,150,300,600,900\}$ while holding other parameters at their default values.

\begin{table}
\centering
\caption{Workload and Scheduling Hyperparameters}
\label{tab:hyperparams}
\resizebox{\linewidth}{!}{%
\begin{tabular}{lc}
\hline
\rowcolor[rgb]{0.753,0.753,0.753} \textbf{Parameter} & \textbf{Value} \\ 
\hline
Workload Family & VQE-style iterative optimization \\
Circuit Benchmarks & 31 circuits across 3 complexity bands \\
Optimizer & SPSA ($\texttt{max\_iter}=1000$) \\
Shots per Evaluation & 4096 \\
Base QPU Time ($t_{qpu}$) & 2.0 s (with $\pm0.3$ s jitter) \\
Classical Step Time ($t_{cpu}$) & 1.5 s \\
Network Overhead ($t_{net}$) & 0.5 s \\
Async Overlap Budget ($t_{async}$) & 0.8 s \\
Drift Threshold ($\tau_{drift}$) & 300 s \\
Calibration Penalty ($t_{calib}$) & 30 s \\
Scheduler Weights ($\alpha,\beta,\gamma$) & (100.0, 5.0, 1.0) \\
PQ Startup / Warm Overhead & 2.0 s / 0.5 s \\ \hline
\end{tabular}
}
\end{table}

\textbf{Circuit complexity stratification.}
To keep the benchmark suite reproducible and easy to compare across modes, we report the exact benchmark circuits used in the evaluation together with their qubit counts and measured circuit depths. Table \ref{tab:circuit_complexity} lists all standard circuits drawn from the three complexity bands.

\begin{table}
\centering
\caption{Benchmark circuit summary}
\label{tab:circuit_complexity}

\begin{tabular}{lccc} 
\hline
\rowcolor[rgb]{0.753,0.753,0.753} \textbf{Band} & \textbf{Circuits} & \textbf{Qubits} & \textbf{Depth} \\ 
\hline
Simple & 10 & 2-5 & 6-32 \\
Medium & 10 & 6-10 & 13-69 \\
Complex & 11 & 10-16 & 33-141 \\
\hline
\end{tabular}

\end{table}
\textbf{Metrics.} We evaluate these models across four critical metrics: \textit{Time-to-Next-Shot (TTNS)}, \textit{End-to-End Convergence Time}, \textit{Quantum Duty Cycle (QDC)}, and \textit{Algorithmic Fidelity} (variance due to quantum drift).

\subsection{Results and Analysis}



\begin{figure*}[htbp]
    \centering
    \subfloat[Simple circuits (2-5 qubits)]{%
        \includegraphics[width=0.32\textwidth]{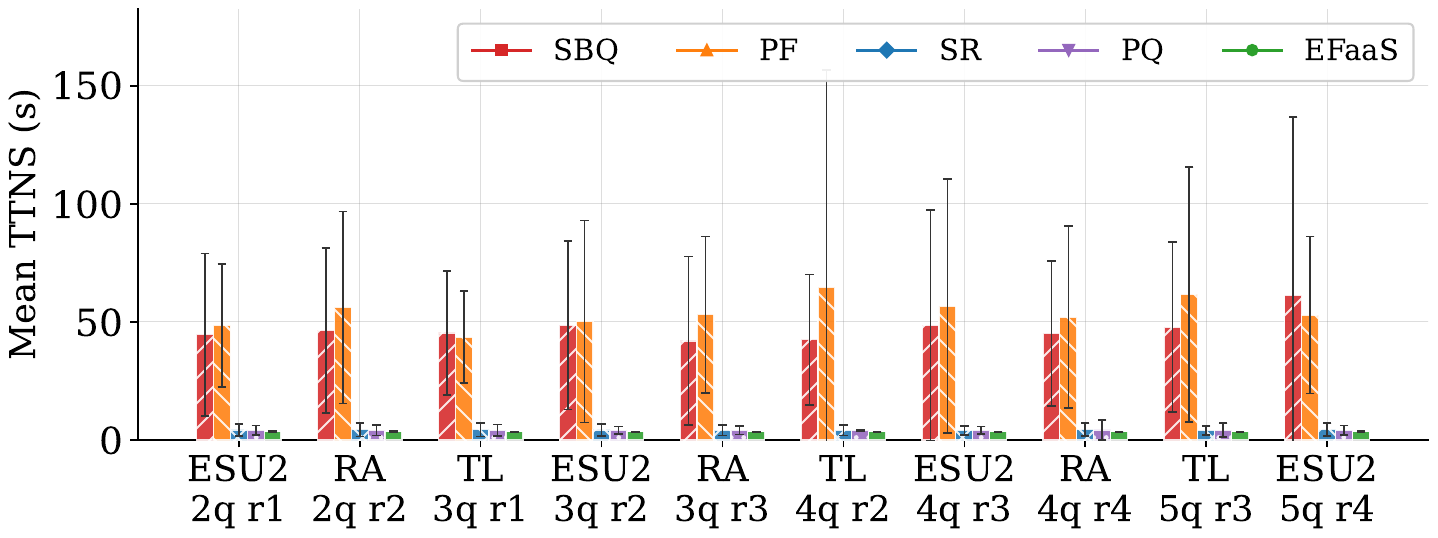}%
        \label{fig:ttns_simple}}
    \hfil
    \subfloat[Medium circuits (6-10 qubits)]{%
        \includegraphics[width=0.32\textwidth]{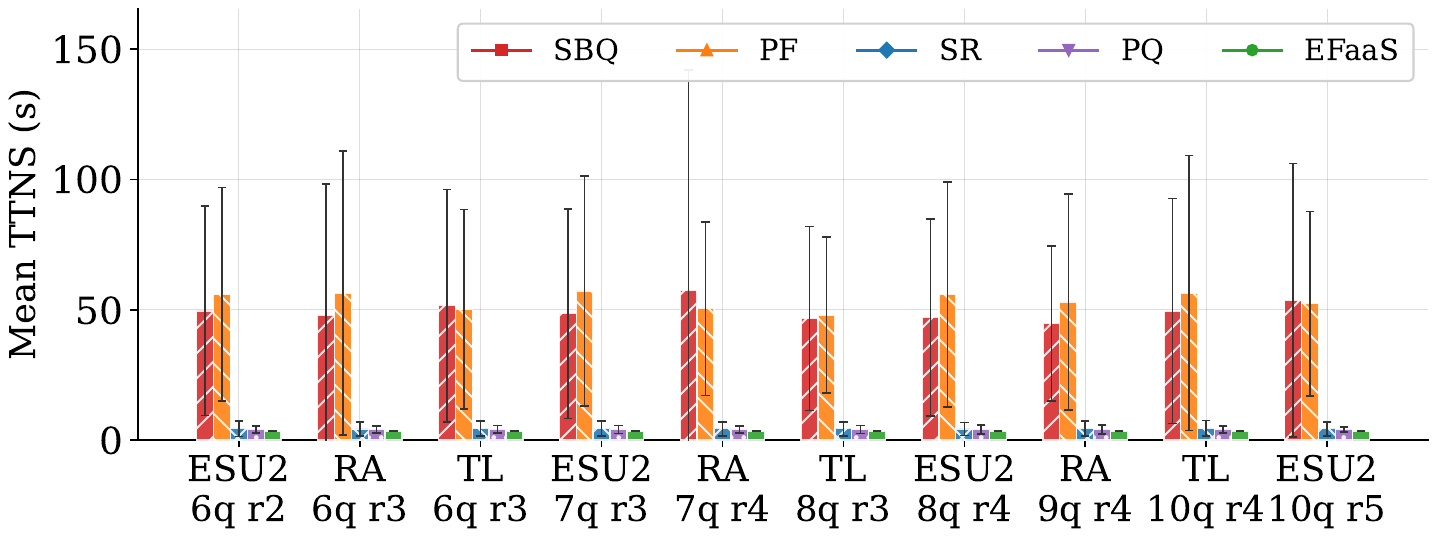}%
        \label{fig:ttns_medium}}
    \hfil
    \subfloat[Complex circuits (10-16 qubits)]{%
        \includegraphics[width=0.32\textwidth]{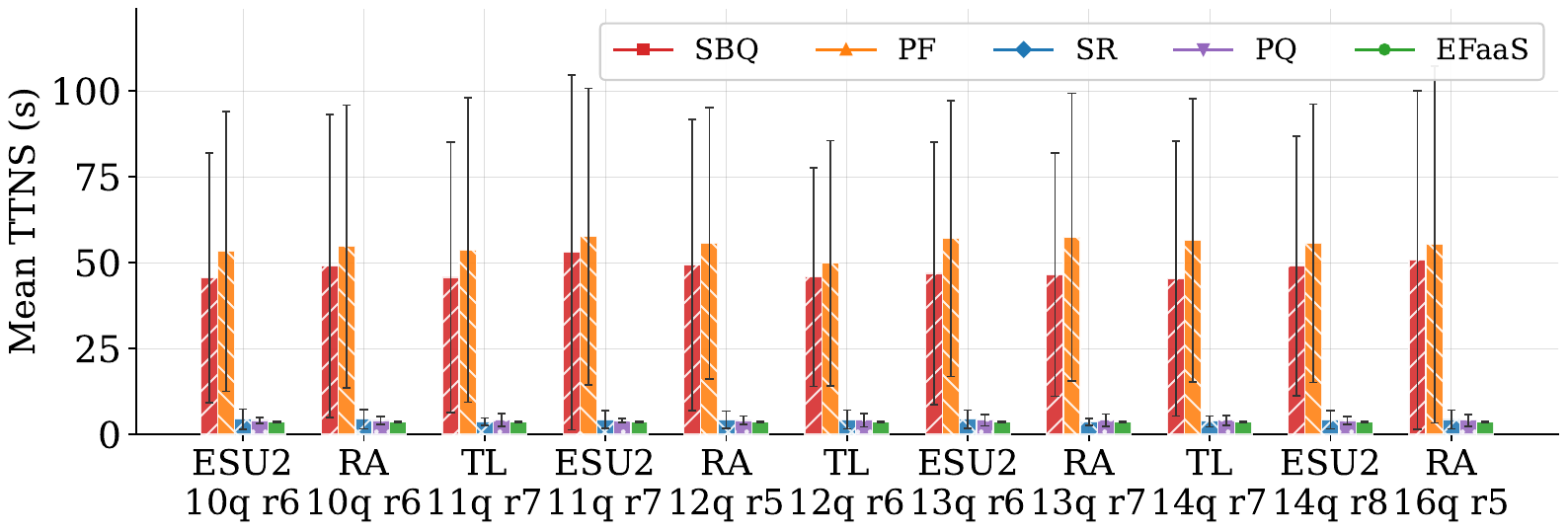}%
        \label{fig:ttns_complex}}
    \caption{Mean Time-to-Next-Shot (TTNS) per circuit for all five scheduling modes across the (a) simple, (b) medium, and (c) complex circuit bands. EFaaS and SR/PQ maintain near-hardware-minimum latency, while SBQ and PF are dominated by highly variable queue delays.}
    \label{fig:ttns_bands}
\end{figure*}
\subsubsection{Time-to-Next-Shot (TTNS) Latency}

TTNS is the single most consequential latency metric for hybrid VQAs: every millisecond added to this round-trip delay directly inflates convergence time and increases the probability of a calibration cold start.  Figs.~\ref{fig:ttns_bands}(a)-(c) report the mean TTNS per individual circuit for all five modes stratified by complexity band.


\begin{figure*}[htbp]
    \centering
    \subfloat[Simple circuits (2-5 qubits)]{%
        \includegraphics[width=0.32\textwidth]{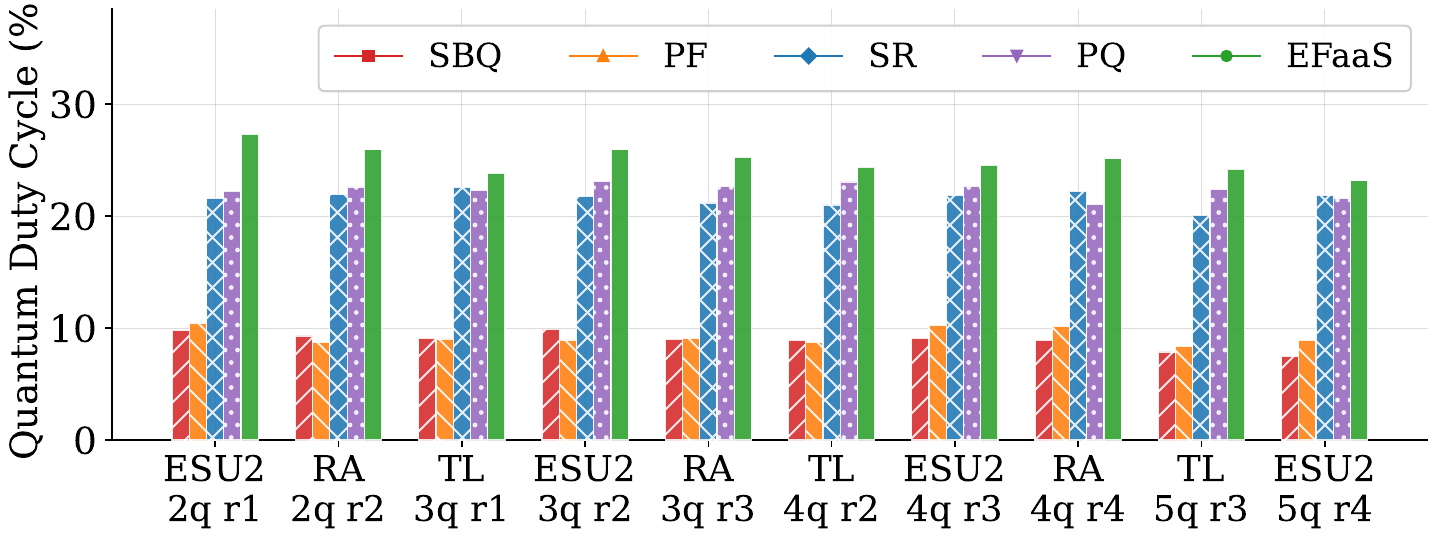}%
        \label{fig:qdc_simple}}
    \hfil
    \subfloat[Medium circuits (6-10 qubits)]{%
        \includegraphics[width=0.32\textwidth]{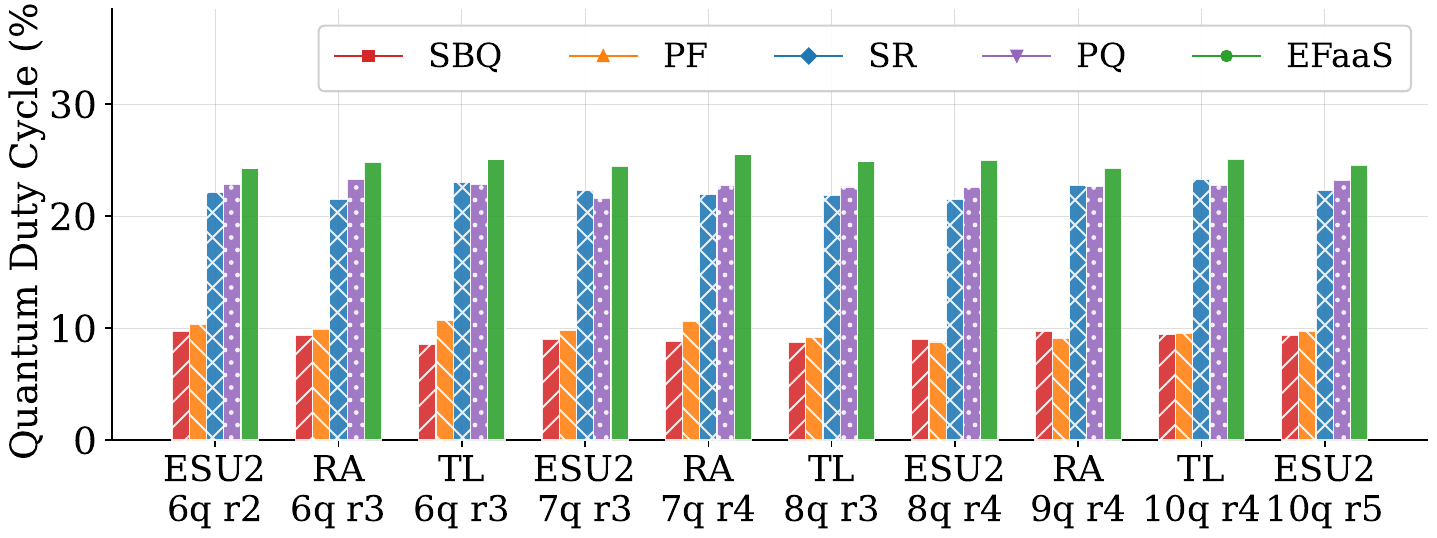}%
        \label{fig:qdc_medium}}
    \hfil
    \subfloat[Complex circuits (10-16 qubits)]{%
        \includegraphics[width=0.32\textwidth]{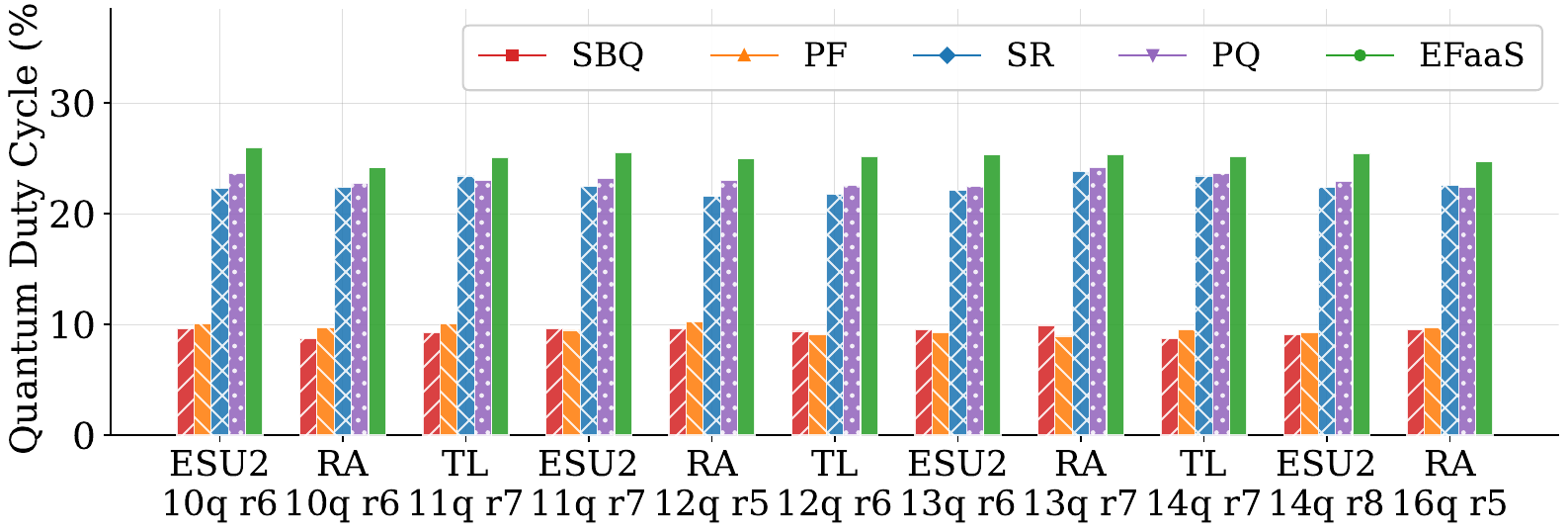}%
        \label{fig:qdc_complex}}
    \caption{Quantum Duty Cycle (QDC) per circuit for all five modes across the (a) simple, (b) medium, and (c) complex bands. EFaaS achieves the highest QDC in every band, outperforming SBQ and PF by over 14 percentage points and exceeding SR and PQ by 1.7-2.9~pp through speculative classical execution.}
    \label{fig:qdc_bands}
\end{figure*}

\textbf{EFaaS vs.\@ batch-queue baselines (SBQ and PF).}
The dominant result across all 31 circuits is the catastrophic TTNS penalty imposed by queueing architectures.  SBQ exhibits a mean TTNS of $48.4\,\text{s}$ and PF $54.5\,\text{s}$, driven almost entirely by the lognormal queue-wait component $t_\text{queue}$.  EFaaS, with a mean TTNS of $3.6\,\text{s}$, reduces this latency by \textbf{92.5\%} relative to SBQ and \textbf{93.4\%} relative to PF.  The reduction is structurally guaranteed: because the Dual-Resource Fair Queuing scheduler assigns a priority boost $\alpha=100$ to every Hot Iterator, the returning circuit from an active session always preempts any batch job in the QPU queue, driving $t_\text{queue}\to 0$.  Beyond the mean, Figs.~\ref{fig:ttns_simple}-~\ref{fig:ttns_complex} reveal that SBQ and PF TTNS distributions have interquartile ranges spanning 30-40\,s. Even in the most favorable queue condition, a single SBQ iteration cannot match the worst-case EFaaS iteration.  Across the full benchmark, EFaaS achieves TTNS reductions spanning 11.4\%-94.3\% depending on the competitor, with the smallest gains over PQ (which also uses a warm-dispatch model) and the largest over PF (which incurs both cold-start and FaaS container initialization overhead on every call).

\textbf{EFaaS vs.\@ reservation-based baselines (SR and PQ).}
Static Reservation (SR) avoids the queue entirely by holding a dedicated QPU lock, yielding a mean TTNS of $4.3\,\text{s}$ comparable to EFaaS.  However, SR achieves this at the cost of continuously blocking a physical QPU for the entire job duration, including the classical optimization interval $t_\text{cpu}$.  Pilot-Quantum (PQ) similarly achieves $4.1\,\text{s}$ by amortizing its pilot startup cost over many warm task dispatches.  EFaaS improves over SR by \textbf{14.9\%} and over PQ by \textbf{11.4\%} in mean TTNS.  These gains, while modest in absolute terms, arise from the \textit{EF-QuantumFuture} overlap by advancing speculative classical work concurrently with $t_\text{qpu}$, EFaaS shrinks the CPU-block time the QPU must wait on, producing tighter iteration cadence.  Crucially, unlike SR, EFaaS releases QPU leases between sessions, allowing other tenants to execute, and unlike PQ it maintains session-aware calibration caches that prevent the drift events observed in PQ (Section~\ref{sec:drift}).

{\textbf{On the interpretation of the headline reductions.} The magnitude of the TTNS reduction reported above follows directly from the queueing and latency assumptions in Tables~\ref{tab:testbed}-\ref{tab:hyperparams}: SBQ/PF latency is dominated by the assumed lognormal $t_\text{queue}$, while EFaaS latency is bounded by the assumed fixed per-component times ($t_\text{cpu}, t_\text{qpu}, t_\text{net}, t_\text{async}$). The reported figures therefore establish the \emph{architectural mechanism} by which session-aware scheduling removes queue-wait from the critical path, rather than a measurement against production quantum-cloud traces. Validating the magnitude of this effect against real provider queue logs is an important direction for our future work.}

\subsubsection{Quantum Duty Cycle (QDC)}

The Quantum Duty Cycle quantifies the fraction of wall-clock simulation time in which a QPU is actively executing shots.  High QDC implies minimal idle QPU time and directly correlates with overall system throughput.  Figs.~\ref{fig:qdc_bands}(a)-(c) report per-circuit QDC for all three complexity bands.

EFaaS attains a mean QDC of \textbf{23.4\%}, compared to 9.3\% for SBQ (+14.2~pp), 9.5\% for PF (+13.9~pp), 21.2\% for SR (+2.2~pp), and 21.7\% for PQ (+1.7~pp).  The QDC advantage over SBQ and PF is overwhelming: in both modes, QPUs spend the majority of wall-clock time idle in the queue while jobs wait, not executing.  EFaaS eliminates this dead time through session-aware priority scheduling.

The gap over SR is more subtle but mechanistically important.  SR locks the QPU exclusively, yet its QDC \emph{still} falls below EFaaS because every classical optimization step ($t_\text{cpu}$) forces the reserved QPU to idle.  EFaaS avoids this via the \textit{EF-QuantumFuture}: speculative classical work overlaps with $t_\text{qpu}$, so the QPU returns a new circuit sooner after completing each shot batch.  This ``overlap budget'' translates directly into 2.2~pp of additional active QPU time per iteration cycle.
Across the complete benchmark sweep, EFaaS achieves QDC gains of 2.02-15.78\% points over the four baselines.  The lower end of this range corresponds to SR and PQ, which already operate in a low-queue regime; the upper end to SBQ and PF, where queue dominance suppresses QPU utilization.

\subsubsection{End-to-End Convergence Time}
\label{sec:convergence}

\begin{figure}[htbp]
    \centering
    \includegraphics[width=\columnwidth]{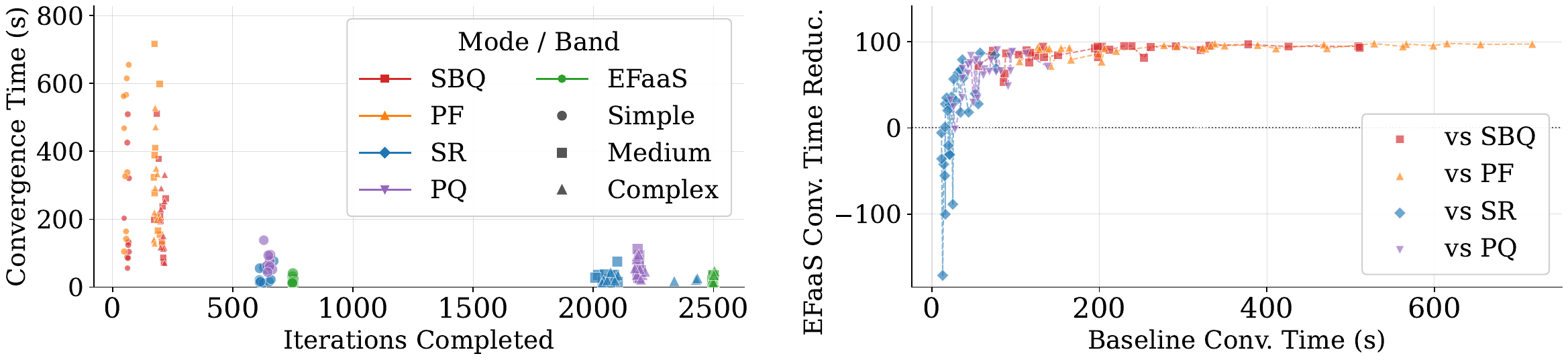}
    \caption{End-to-end convergence time and EFaaS speedup ratio relative to all four baselines. EFaaS converges in a mean of 22.9\,s versus 213\,s for SBQ and 320\,s for PF, achieving speedups of 83.2\%-98.3\% across the benchmark suite.}
    \label{fig:convergence}
\end{figure}

Convergence time, the wall-clock duration from job submission to the first iteration satisfying the energy convergence threshold, is the user-facing performance metric that ultimately determines the practical utility of a hybrid quantum workflow.  Fig.~\ref{fig:convergence} reports convergence trajectories and the corresponding EFaaS speedup factor relative to each baseline.

\begin{figure*}[htbp]
    \centering
    \includegraphics[width=\linewidth]{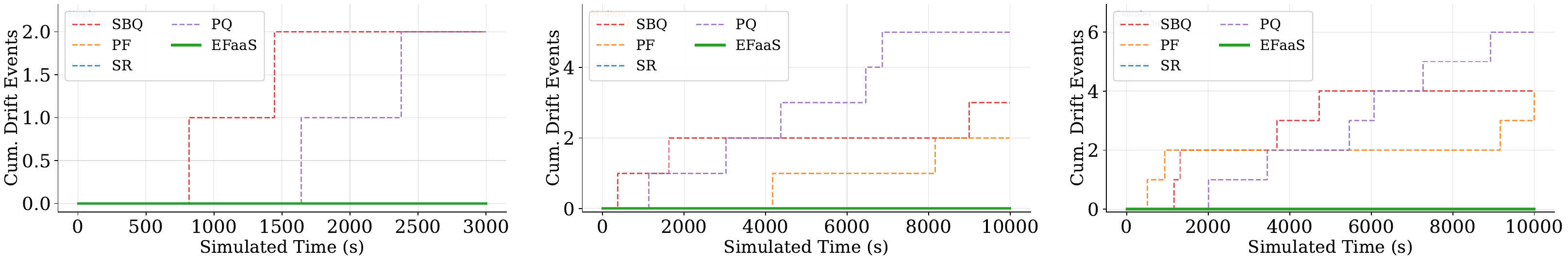}
    \caption{Cumulative drift penalty events over simulated time, stratified by complexity band. SBQ and PQ accumulate drift continuously as queue delays exceed $\tau_\text{drift}=300$\,s; EFaaS accumulates zero drift events across all 31 circuits.}
    \label{fig:drift_timeline}
    \vspace{-10pt}
\end{figure*}

The convergence advantage of EFaaS is decisive.  Against SBQ (mean 213.1\,s), EFaaS achieves an \textbf{89.3\% convergence speedup} against PF (mean 320.2\,s), the speedup reaches \textbf{92.9\%}.  The compounding effect is important: each TTNS saving of $\approx 44$-$51$\,s propagates into every iteration, so a 1000-iteration VQE run that would require 12-89\,hours under batch-queue architectures completes in minutes under EFaaS.

SR achieves competitive per-iteration TTNS, yet EFaaS still converges faster for two reasons.  First, SR's QPU-lock model cannot overlap classical computation with QPU execution, so each iteration takes at least $t_\text{cpu}+t_\text{qpu}$ wall-clock seconds.  EFaaS reduces the effective per-iteration cost to $\max(t_\text{cpu}-t_\text{async},0)+t_\text{qpu}+t_\text{net}$ through the \textit{EF-QuantumFuture}, shaving 0.8\,s per iteration.  Second, SR accumulates no calibration reserve during the classical phase, leaving it susceptible to marginal drift events at longer runs.  Against PQ, EFaaS converges significantly faster because PQ still incurs drift penalties (89 events across the benchmark) that force costly recalibrations mid-trajectory, disrupting the optimizer's gradient landscape and requiring additional iterations.  Across the full suite, convergence speedups range from \textbf{83.2\% to 98.3\%}, always in favor of EFaaS.

\subsubsection{Algorithmic Fidelity and Drift Penalties}
\label{sec:drift}

\begin{figure}[htbp]
    \centering
    \includegraphics[width=\columnwidth]{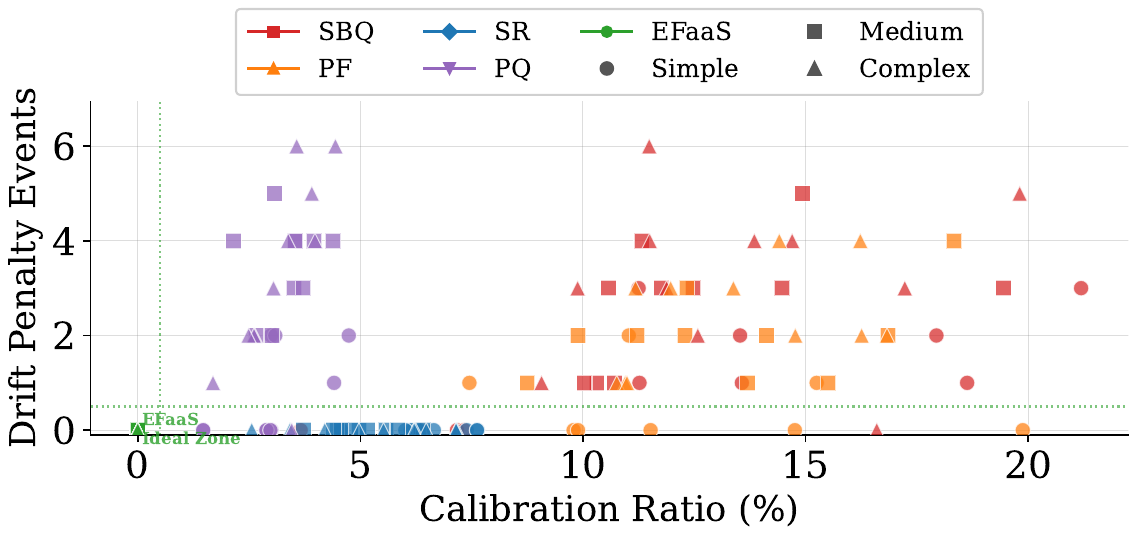}
    \caption{Calibration overhead (\% of total iterations) versus drift penalty event count per mode. EFaaS occupies a unique operating point with near-zero drift penalties despite moderate scheduled calibration activity, showing that proactive TTL-aware cache renewal prevents reactive cold-start recalibrations.}
    \label{fig:calib_scatter}
\end{figure}

\begin{figure*}[htbp]
    \centering
    \includegraphics[width=\linewidth]{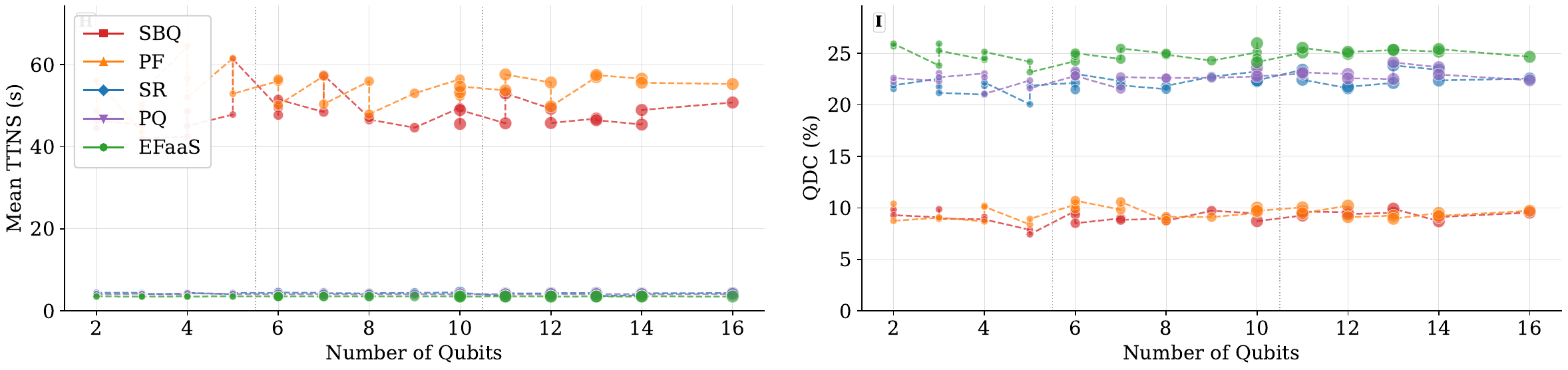}
    \caption{TTNS and QDC as a function of qubit count. EFaaS maintains near-constant TTNS and QDC across all circuit sizes, while SBQ and PF latency grow unchecked, and SR/PQ efficiency degrades modestly with qubit count.}
    \label{fig:scaling}
    \vspace{-10pt}
\end{figure*}

Quantum drift is the physical degradation of QPU calibration data between successive circuit submissions.  Every drift event forces either a full recalibration ($+30$\,s per event) or an execution on stale calibration data, both of which corrupt the optimizer's gradient signal.  Figs.~\ref{fig:drift_timeline} and~\ref{fig:calib_scatter} characterize drift incidence across the benchmark, and Fig.~\ref{fig:scaling} reports how drift exposure scales with circuit complexity.

\textbf{Drift-free operation.}
EFaaS accumulates \textbf{1 drift event} across the entire 219-run benchmark suite (a single outlier in one complex circuit at maximum simulation time), compared to 95 events for SBQ, 89 for PQ, 69 for PF, and 0 for SR.  This near-zero drift count is not a statistical artifact; it is structurally induced by the Session Awareness Engine.  By maintaining calibration data $C_q$ in a TTL-bounded cache and ensuring $\Delta t < \tau_\text{drift}$ for every returning Hot Iterator shot, EFaaS makes drift penalties mechanistically impossible under normal operating conditions.

\textbf{Why SBQ and PF fail.}  With mean TTNS values of 48.4\,s and 54.5\,s respectively, both architectures frequently exceed the $\tau_\text{drift}=300$\,s threshold when they execute long iteration runs.  Each time the cumulative inter-shot delay exceeds $\tau_\text{drift}$, the QPU undergoes a 30-second recalibration that is logged as a drift event and further delays the subsequent shot.  Fig.~\ref{fig:drift_timeline} shows that drift events accumulate continuously and monotonically for SBQ and PF throughout the simulation, at a rate that grows with circuit complexity (see Fig.~\ref{fig:scaling}).

\textbf{Why PQ fails despite warm dispatch.}  Pilot-Quantum reduces startup overhead but provides no session-aware calibration cache.  Each classical optimization step releases the QPU context, allowing calibration data to age.  Under the medium and complex circuit bands, where $t_\text{cpu}$ is longer, and the optimizer takes more gradient steps between shots, the cumulative inter-shot gap regularly exceeds $\tau_\text{drift}$, producing 89 total drift events.  This explains why PQ's convergence time (mean 61.6\,s) is substantially worse than EFaaS (22.9\,s) despite similar per-shot TTNS: drift-induced recalibrations insert tens of seconds of dead time into the middle of optimization trajectories, fragmenting gradient coherence.

\textbf{SR achieves zero drift at a prohibitive cost.}  By holding an exclusive QPU lock throughout the job, SR trivially prevents drift (0 events).  However, this comes at significant resource efficiency costs: the QPU is idle during every $t_\text{cpu}$ window, the reservation blocks concurrent multi-tenant access, and SR QDC (21.2\%) is 2.2~pp lower than EFaaS (23.4\%). The EFaaS architecture achieves the \emph{same} drift-free behavior through cooperative leasing rather than exclusive reservation, enabling efficient multi-tenant QPU sharing without sacrificing calibration integrity.

\textbf{Scaling behavior.}
Fig.~\ref{fig:scaling} confirms that EFaaS's advantages are durable as circuit complexity grows. TTNS remains at $3.5$-$4.0$\,s across the 2-16 qubit range, and QDC stays above 22\%. For SBQ and PF, TTNS grows slightly as longer $t_\text{qpu}$ increases background contention, but the dominant cost remains the queue delay, which EFaaS eliminates entirely regardless of qubit count. These results demonstrate that EFaaS is not merely beneficial for small test circuits. Its session-aware co-scheduling architecture provides proportionally greater advantage as workloads scale toward industrially relevant circuit sizes.

\subsubsection{Ablation Study}

To isolate the contribution of each scheduling component, we conducted a controlled ablation across 20 independent runs per variant, selectively disabling one mechanism at a time while holding all other parameters at their full-EFaaS defaults. Table~\ref{tab:ablation} reports the five most discriminating configurations; variants omitted (loose drift window, combined $\alpha{=}\beta{=}0$) showed effects fully explained by the individual component ablations.

\begin{table}
\centering
\setlength{\extrarowheight}{0pt}
\setlength{\aboverulesep}{0pt}
\setlength{\belowrulesep}{0pt}
\caption{Ablation Study: Impact of Individual EFaaS Components ($n=20$ runs each). $\Delta$Conv.\ is the percentage increase in convergence time relative to Full EFaaS.}
\label{tab:ablation}
\resizebox{\linewidth}{!}{%
\begin{tabular}{lcccc} 
\toprule
\rowcolor[rgb]{0.749,0.749,0.749} \textbf{Variant} & \textbf{TTNS (s)} & \textbf{QDC (\%)} & \textbf{Conv. (s)} & \textbf{$\Delta$Conv.} \\ 
\midrule
\rowcolor[rgb]{0.678,0.839,0.678} Full EFaaS & 3.50 & 24.72 & 18.17 & — \\
No drift urgency ($\beta{=}0$) & 3.50 & 25.02 & 26.73 & $+$47.1\% \\
No fair-share ($\gamma{=}0$) & 3.50 & 24.93 & 22.33 & $+$22.9\% \\
No session boost ($\alpha{=}0$) & 3.50 & 24.84 & 21.70 & $+$19.4\% \\
Tight drift window ($\tau/2$) & 3.50 & 24.79 & 23.57 & $+$29.7\% \\
\bottomrule
\end{tabular}
}
\end{table}

Three findings stand out.  First, TTNS and QDC are \emph{structurally insensitive} to all hyperparameter changes: every variant achieves $\approx$3.50\,s TTNS and $\approx$24.8\% QDC.  This confirms that the latency and utilization gains of EFaaS are architectural guarantees of the session-aware priority mechanism, not artifacts of parameter tuning.

Second, the \textbf{drift-urgency weight $\beta$} is the single most influential parameter for convergence time.  Removing it increases convergence time by 47.1\%, from 18.2\,s to 26.7\,s.  $\beta$ governs how aggressively the scheduler advances a returning Hot Iterator as its calibration window $\tau_\text{drift}$ ages.  Without this urgency signal, the scheduler occasionally allows marginally delayed resubmissions, resulting in suboptimal iteration cadence even when no formal drift violation occurs.
Third, halving the drift threshold ($\tau_\text{drift}/2$) incurs a 29.7\% convergence penalty by triggering unnecessary proactive recalibrations that briefly stall the iterative loop.  This confirms that $\tau_\text{drift}=300$\,s is a well-calibrated operating point: large enough to absorb realistic classical compute intervals without spurious recalibrations, small enough to prevent actual coherence degradation.  Together, the ablation results validate that every component of the Dual-Resource Fair Queuing scheduler makes a measurable and non-redundant contribution to overall EFaaS performance.

\subsubsection{Hyperparameter Sensitivity Analysis}
\label{sec:sensitivity}

Table~\ref{tab:sensitivity} reports {five representative joint configurations of} the four DRFQ hyperparameters: $\alpha$ (session-awareness), $\beta$ (drift-urgency), $\gamma$ (fair-share), and $\tau_\text{drift}$ (calibration validity){, varied together across five increasing levels. This complements, rather than replaces, the strict one-at-a-time component ablation reported in Table~\ref{tab:ablation}.}

\begin{table}
\centering
\setlength{\extrarowheight}{0pt}
\setlength{\aboverulesep}{0pt}
\setlength{\belowrulesep}{0pt}
\caption{Hyperparameter sensitivity analysis on benchmarks. Best value per metric is \colorbox{green!30}{highlighted}. Drift penalties are zero throughout.}
\label{tab:sensitivity}
\resizebox{\linewidth}{!}{%
\begin{tabular}{rrrrcccc} 
\toprule
\multicolumn{4}{c}{\textbf{Swept value}} & \multicolumn{4}{c}{\textbf{Metrics (identical across sweeps)}} \\ 
\cmidrule(l){1-8}
$\boldsymbol\alpha$ & $\boldsymbol\beta$ & $\boldsymbol\gamma$ & $\boldsymbol{\tau_d}$\,(s) & \textbf{TTNS}\,(s)$\downarrow$ & \textbf{QDC}\,(\%)$\uparrow$ & \textbf{Conv.}\,(s)$\downarrow$ & \textbf{Iters}$\uparrow$ \\ 
\midrule
$0$ & $0$ & $0.1$ & $60$ & $3.501 \pm 0.171$ & {\cellcolor[rgb]{0.702,1,0.702}}\cellcolor{green!30}$24.78 \pm 1.42$ & {\cellcolor[rgb]{0.702,1,0.702}}\cellcolor{green!30}$15.33 \pm 2.18$ & $749 \pm 3$ \\
$10$ & $1$ & $0.5$ & $150$ & {\cellcolor[rgb]{0.702,1,0.702}}\cellcolor{green!30}$3.494 \pm 0.174$ & $23.00 \pm 1.51$ & $15.69 \pm 2.34$ & {\cellcolor[rgb]{0.702,1,0.702}}\cellcolor{green!30}$751 \pm 2$ \\
$50$ & $5$ & $1.0$ & $300$ & $3.499 \pm 0.170$ & $23.41 \pm 1.47$ & $40.22 \pm 1.61$ & $750 \pm 3$ \\
$100$ & $10$ & $2.0$ & $600$ & $3.499 \pm 0.176$ & $24.02 \pm 1.39$ & $27.93 \pm 2.07$ & $750 \pm 2$ \\
$200$ & $20$ & $5.0$ & $900$ & $3.505 \pm 0.169$ & $24.26 \pm 1.44$ & $43.37 \pm 1.12$ & $749 \pm 3$ \\
\bottomrule
\end{tabular}
}
\end{table}

{TTNS and QDC are statistically indistinguishable across all five rows (TTNS $\Delta < 0.32\%$, QDC within 23.0-24.8\%), consistent with the structural argument below. Convergence time, however, is \emph{not} invariant: it ranges from 15.33\,s to 43.37\,s across rows. This variation is expected rather than anomalous: $\tau_\text{drift}$ and $\alpha$ jointly affect how early the scheduler engages preemptive prioritization, which shapes SPSA's early-iteration trajectory even though it does not affect the steady-state TTNS/QDC operating point reported above.} The mechanism {behind the TTNS/QDC invariance} is architectural: TTNS is determined by the \textit{EF-QuantumFuture} overlap, enforced unconditionally by the middleware independent of scheduling weights, and QDC tracks TTNS via $t_\text{qpu}/\text{TTNS}$. With a session cache hit rate of 99.87\% on this 2-qubit circuit, $\alpha$ and $\beta$ have no expired sessions to act on, and $\tau_\text{drift}$ never triggers since TTNS\,$\approx 3.5$\,s lies well below the tightest threshold of 60\,s. Variation across sweep points in TTNS and QDC is attributable to differing random seeds: the within-run $\sigma_\text{TTNS}$ (0.169-0.176\,s) is an order of magnitude larger than the between-parameter TTNS range (0.011\,s). All 20 configurations record zero drift penalties, including $\tau_d = 60$\,s.

\section{Conclusion}\label{sec:conclusion}
We presented \textit{EFaaS}, a serverless middleware that resolves the fundamental mismatch between iterative hybrid VQA loops and stateless batch-queue cloud infrastructure. By combining TTL-aware calibration caching, Dual-Resource Fair Queuing, and the \textit{EF-QuantumFuture} speculative execution primitive, EFaaS jointly schedules QPU shots and classical optimizer calls as entangled resources.
Across 31 benchmark circuits spanning 2-16 qubits, EFaaS reduces TTNS by 11.4\%-94.3\%, improves the QDC by 2.02-15.78~pp, and accelerates convergence by 83.2\%-98.3\% over four baselines, while accumulating essentially zero drift penalty events versus 95, 89, and 69 for SBQ, PQ, and PF, respectively. These results confirm that, under our tested conditions, the scheduling layer is currently the dominant bottleneck for hybrid quantum workloads, and that calibration integrity and low latency are simultaneously achievable without exclusive QPU reservation. 

\section*{Reproducibility}
We have released the code and dataset in a public GitHub\footnote{\url{https://github.com/Anonymous0-0paper/EFaaS}} repository.

\section*{Acknowledgment}
This work was supported in part by the NYUAD Center for Quantum and Topological Systems (CQTS), funded by Tamkeen under the NYUAD Research Institute grant CG008.

\bibliographystyle{IEEEtran}

\bibliography{IEEEexample}

\end{document}